\documentclass[11pt]{article}
\usepackage{authblk}
\usepackage{amsmath, amssymb, graphics, setspace}
\usepackage[pdftex]{color,graphicx}
\usepackage[superscript,nomove,nosort]{cite} %
\usepackage{verbatim}
\usepackage{multicol}
\usepackage[margin=1.0in]{geometry}
\usepackage{braket}
\begin{document}
\title{Quantum entanglement between an atom and a molecule}
\author[1,2,4,5]{Yiheng Lin}
\author[2,3]{David R. Leibrandt}
\author[2,3]{Dietrich Leibfried}
\author[2]{Chin-wen Chou}
\affil[1]{CAS Key Laboratory of Microscale Magnetic Resonance and Department of Modern Physics, University of Science and Technology of China, Hefei 230026, China}
\affil[2]{Time and Frequency Division, National Institute of Standards and Technology, 325 Broadway, Boulder, CO 80305, USA}
\affil[3]{Department of Physics , University of Colorado, Boulder, Colorado, USA.}
\affil[4]{Hefei National Laboratory for Physical Sciences at the Microscale, University of Science and Technology of China, Hefei 230026, China
}
\affil[5]{Synergetic Innovation Center of Quantum Information and Quantum Physics, University of Science and Technology of China, Hefei 230026, China
}
\date{\today}
\maketitle
Conventional information processors freely convert information between different physical carriers to process, store, or transmit information. It seems plausible that quantum information will also be held by different physical carriers in applications such as tests of fundamental physics, quantum-enhanced sensors, and quantum information processing. Quantum-controlled molecules in particular could transduce quantum information across a wide range of quantum-bit (qubit) frequencies, from a few kHz for transitions within the same rotational manifold \cite{chou_preparation_2017}, a few GHz for hyperfine transitions, up to a few THz for rotational transitions, to hundreds of THz for fundamental and overtone vibrational and electronic transitions, possibly all within the same molecule. Here, we report the first demonstration of entanglement between states of the rotation of a $\rm^{40}CaH^+$ molecular ion and internal states of a $\rm^{40}Ca^+$ atomic ion \cite{ref40Ca+}. The qubit addressed in the molecule has a frequency of either 13.4~kHz \cite{chou_preparation_2017} or 855~GHz \cite{choucomb2019}, highlighting the versatility of molecular qubits. This work demonstrates how molecules can transduce quantum information between qubits with different frequencies to enable hybrid quantum systems. We anticipate that quantum control and measurement of molecules as demonstrated here will create opportunities for quantum information science, quantum sensors, fundamental and applied physics, and controlled quantum chemistry.\\
\\

Quantum state control of atoms has enabled high-fidelity entangling gates for large-scale quantum computation \cite{ballance_high-fidelity_2016,gaebler_high-fidelity_2016,levine_high-fidelity_2018}, quantum simulations  \cite{bernien_probing_2017,zhang_observation_2017}, and multi-partite entanglement generation \cite{mcconnell_entanglement_2015,bohnet_quantum_2016,luo_deterministic_2017,omran_generation_2019}. By adding vibrational and rotational degrees of freedom, as well as coupling of multiple angular momenta to the internal state structure, molecules offer unique opportunities in quantum information processing \cite{demille_quantum_2002}, precision measurements \cite{kozyryev_precision_2017,cairncross_precision_2017,altuntas_demonstration_2018,acme_collaboration_improved_2018}, and tests of fundamental physics \cite{safronova_search_2018}. Following ideas inspired by laser cooling, trapping, and quantum state control of atoms, quantum control of molecules has made substantial progress. Cold atoms have been associated to produce cold molecular ensembles and single molecules \cite{moses_new_2017, liu_building_2018}. Sub-Doppler laser cooling of the translational motion \cite{truppe_molecules_2017} and initialization of rotational and vibrational states of molecules have been demonstrated \cite{ospelkaus_controlling_2010,reinaudi_optical_2012,park_ultracold_2015}, dipolar and chemical interactions between molecules have been explored \cite{yan_observation_2013, hu_direct_2019, cheuk_observation_2020}, as well as resonant atom-molecule cold collisions \cite{yang_observation_2019}. For trapped molecular ions, precision spectroscopy of rotational and vibrational energy levels has been demonstrated \cite{biesheuvel_probing_2016,alighanbari_rotational_2018} and quantum-logic spectroscopy (QLS)\cite{schmidt_spectroscopy_2005} has been introduced as an alternative to techniques pioneered on neutral atoms for state detection \cite{wolf_non-destructive_2016,sinhal_quantum_2019}, preparation, and manipulation of molecular ions \cite{chou_preparation_2017}. \\

Towards fully realizing the potential applications of molecules in quantum science, demonstration of entanglement involving an individually controlled molecule is a necessary and critical step. Recent proposals suggest using neutral molecules and molecular ions to form qubits \cite{demille_quantum_2002,hudson_dipolar_2018-1,ni_dipolar_2018}, and explore their electric dipole moments for long-range interactions. Molecules can also facilitate the construction of a hybrid quantum system. For example, molecules with permanent electric dipole moments can serve as antennas for coupling to quantum systems of disparate nature at very different frequencies, including mechanical cantilevers \cite{kippenberg_cavity_2008} and microwave photons in a superconducting cavity \cite{schuster_cavity_2011}. For coupled atomic-molecular ion systems \cite{campbell_dipole-phonon_2019}, an atomic ion can also function as a means to prepare entangled states of several co-trapped molecular ions that can be used for quantum-enhanced metrology and sensing over the wide frequency range covered by molecules. If control based on QLS can be extended to vibrational transitions and their overtones, a molecular ion can serve as a bridge to connect atomic ion qubits to many other systems, for example electromagnetic radiation with frequency up to several hundred THz, which includes low-transmission-loss photonic (flying) qubits in the 1.5~$\mu$m to 1.6~$\mu$m telecom wavelength range.\\

%
Here we leverage QLS to entangle different rotational levels of a molecular ion with an atomic ion, extending previous work \cite{chou_preparation_2017,choucomb2019} on molecular state preparation, detection, and single qubit control. We trap a single molecular ion alongside a well-controlled atomic ion. The atom is utilized for laser cooling of the coupled translational harmonic motion of the two ions, for preparation of a pure quantum state of the molecule \cite{chou_preparation_2017}, and to serve as a high-fidelity qubit \cite{ref40Ca+} in our entanglement demonstration. We then apply tailored laser pulse sequences to generate entangled states of the rotation of the molecular ion and the long-lived 411.0~THz (729~nm) electronic qubit of the atomic ion. We achieve an entangled state fidelity of 0.87(3) for a molecular qubit within a rotational manifold 
with a frequency of $\approx$ 13.4~kHz, and a fidelity of 0.76(3) for a qubit in different rotational manifolds with a much higher frequency 
of approximately 855~GHz. In both cases, we obtain fidelities far above the threshold for genuine two-partite entanglement of $1/2$ (ref.\cite{leibfried_creation_2005}). Our demonstration provides evidence that atoms can be entangled with various molecular rotational states, with their frequency differences spanning more than seven orders of magnitude.  \\

\begin{figure}
\centering
\includegraphics[width=\textwidth]{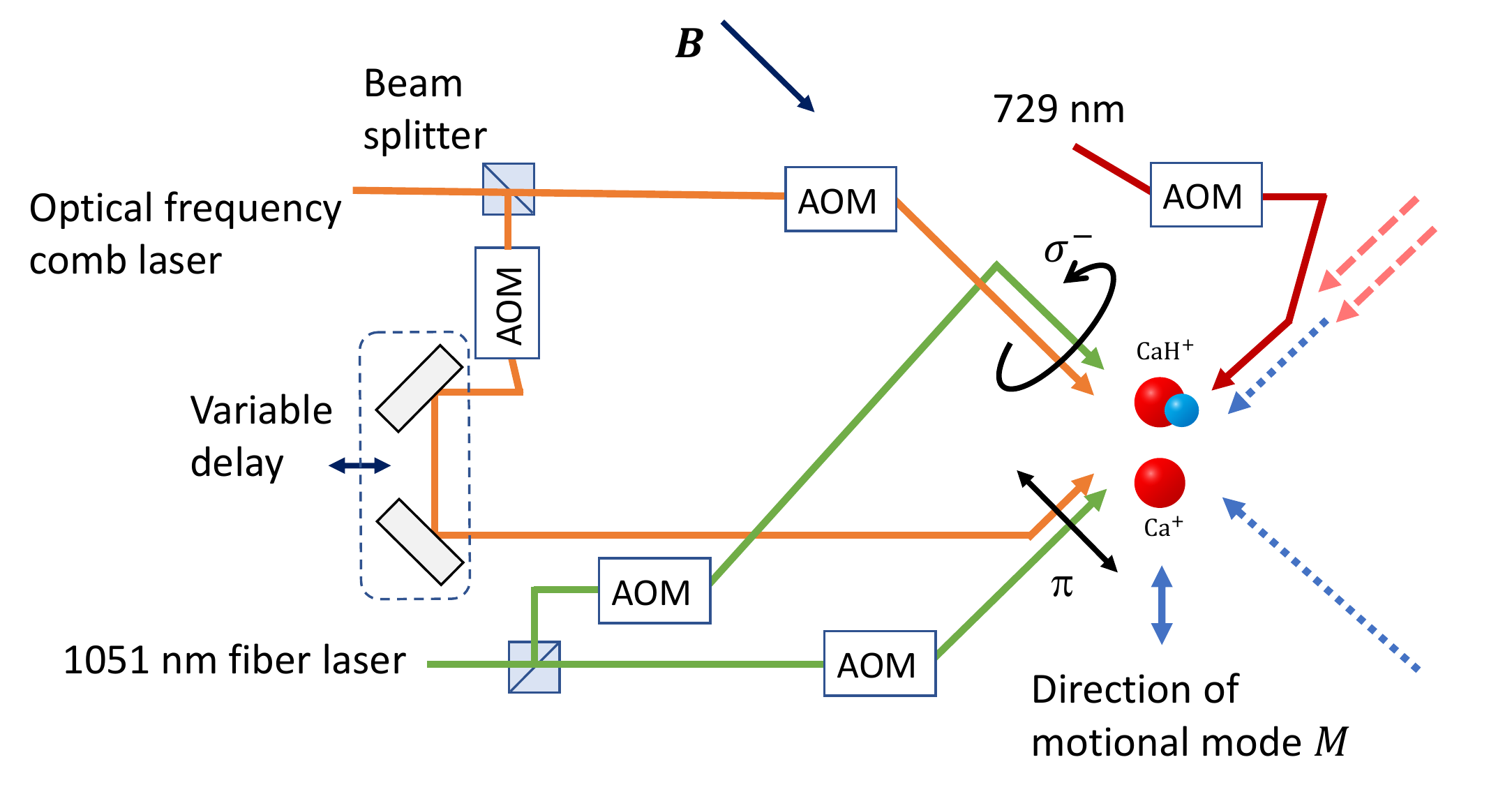}
\caption{
\label{fig:LasBea}
Schematic of the experiment. Acousto-optic modulators (AOMs) are used to control the intensities, frequencies, and relative phases of the light fields on the ions. Transitions between the $\ket{S}$ and $\ket{D}$ states of the $^{40}\rm{Ca}^+$ ion are driven by a laser at 729~nm (red solid line). Other laser beams on the right with dashed arrows are used for cooling, manipulation, and detection of the $^{40}\rm{Ca}^+$ ion (see Methods). The molecule is manipulated by driving two-photon stimulated Raman transitions with two beams generated from a continuous wave 1051 nm fiber laser (green lines), and two beams generated from an optical frequency comb laser centered at $\approx 850$~nm (orange lines). The pairs of beams are offset in frequency by two AOMs and adjusted to circular ($\sigma^-$) and linear ($\pi$) polarization, respectively. The $\sigma^-$ beams are along the direction of the magnetic field ($\boldsymbol B$), which is $\approx$ 45 degrees from the line connecting the equilibrium locations of the ions. The variable delay is adjusted to ensure that the $\approx 40$~fs pulses of the two beams from the frequency comb overlap temporally on the molecule. The direction of the motional mode $M$ is along the ion equilibrium positions as depicted.}
\end{figure}
\begin{figure}
\centering
\includegraphics[width=\textwidth]{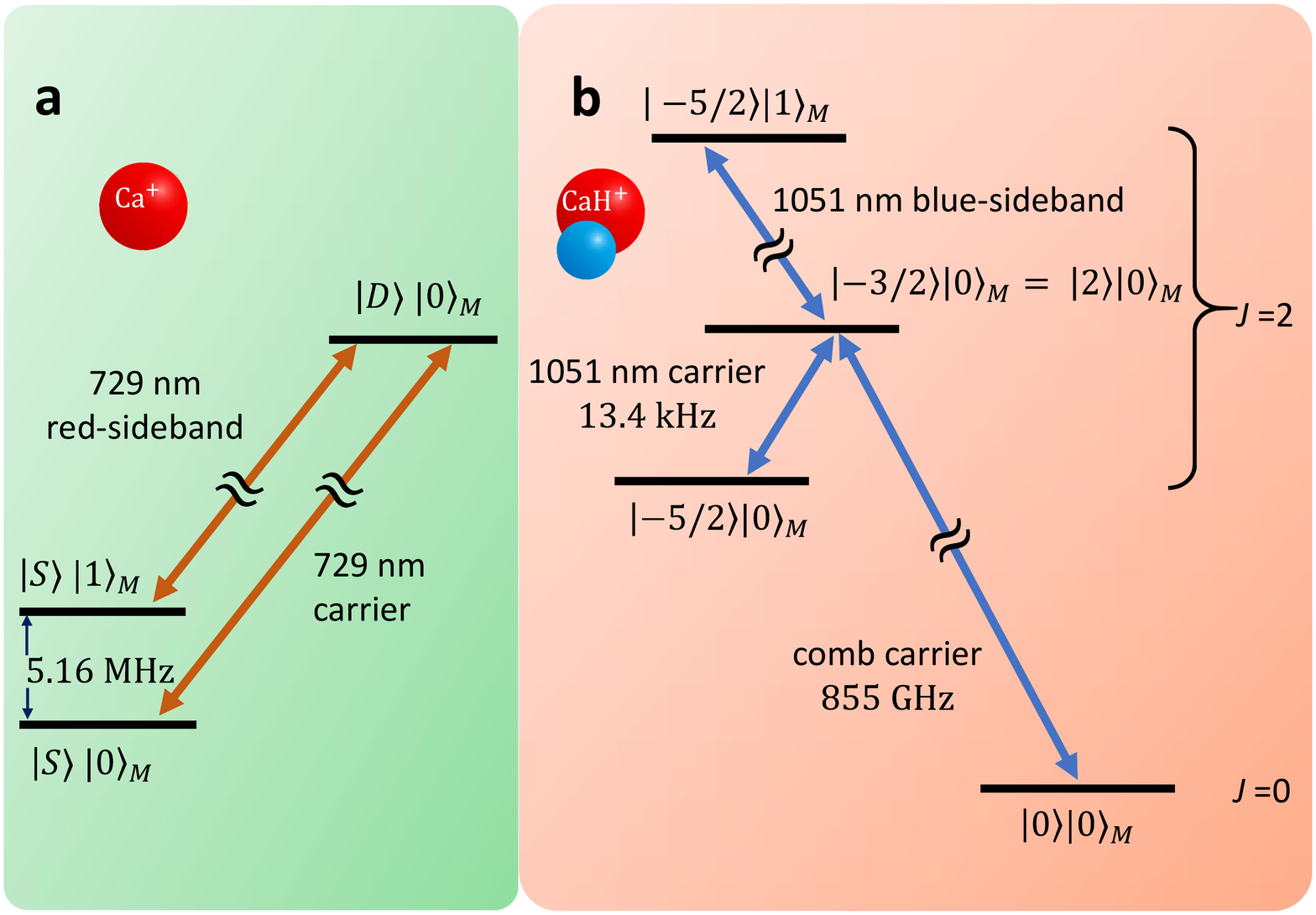}
\caption{\label{fig:EneLev} Energy levels and selected laser-driven transitions for the $\rm ^{40}Ca^+$ atomic ion ({\bf a}) and the $^{40}$CaH$^+$ molecular ion ({\bf b}). Here ``carrier'' denotes transitions between ion states not changing the state $\ket{n}_M$ of the out-of-phase motional mode $M$, while ``sideband'' transitions add or subtract one motional quantum for this mode along with changing the ion state. As described in more detail in the main text, $\ket{S}$ and $\ket{D}$ are electronic states of the atom, and $\ket{-5/2}$, $\ket{-3/2}=\ket{2}$, and $\ket{0}$ denote rotational states of the molecule.}
\end{figure}

In our experiments, a $^{40}\rm Ca^+$ atomic ion is co-trapped with a $^{40}\rm CaH^+$ molecular ion in a linear Paul trap \cite{chou_preparation_2017} (see Fig.~\ref{fig:LasBea}). A static external magnetic field $\boldsymbol B$ with magnitude $\approx 0.36$~mT provides a quantization axis. The Coulomb repulsion between the ions results in two normal modes of coupled harmonic motion along each of three orthogonal directions, with the two ions moving in phase or out of phase. These modes of coupled motion are cooled by applying lasers that are all near resonant with transitions in the $\rm ^{40} Ca^+$ ion (see Methods). To transfer and manipulate quantum states of the co-trapped molecular ion by QLS~\cite{schmidt_spectroscopy_2005}, we utilize the out-of-phase mode $M$ at~$\approx 5.16$~MHz along the direction linking the equilibrium positions of the ions. The quantized state with $n$ phonons (motional quanta) of this mode is denoted by $\ket{n}_M$. For ground state cooling of the coupled motional modes and preparation of entangled states, in the $^{40}$Ca$^+$ atom we use the ground electronic state $\ket{S}\equiv\ket{S_{1/2},m_j=-1/2}$ and a meta-stable excited state $\ket{D}\equiv\ket{D_{5/2},m_j=-5/2}$ with a lifetime of approximately 1 s (see Fig.~\ref{fig:EneLev}\textbf{a}), where $m_j$ is the quantum number for the component of the electronic angular momentum along $\boldsymbol B$. These atomic qubit states are coupled by driving an electric quadrupole transition with a laser near 729 nm (see Fig.~\ref{fig:LasBea}) and can be distinguished by state-dependent fluorescence detection (see Methods). 
\\

The molecule is in its $\Sigma$ electronic and vibrational ground state at room temperature. The rotational eigenstates in the presence of $\boldsymbol{B}$ are denoted as $\ket{J,m,\xi}$ , where the non-negative integer $J$ is the rotational angular momentum quantum number, $m=m_J+m_I$ is the sum of the components of the rotational angular momentum ($-J\leq m_J\leq J $) and the proton spin ($m_I=\pm1/2$) along $\boldsymbol B$, and $\xi=\{+,-\}$ distinguishes the two eigenstates with the same $m$ that are split in energy due to the interaction between the rotational angular momentum and nuclear spin, except for the stretch states with $m=\pm(J+1/2)$ where $\xi$ indicates the sign of $m$ (ref.\cite{chou_preparation_2017}). States within a rotational manifold with the same $J$ are manipulated by driving stimulated Raman transitions with light fields from a fiber laser at 1051~nm (ref.\cite{chou_preparation_2017}) as schematically shown in Fig.~\ref{fig:LasBea}. In particular, we can tune the frequency difference between these light fields to address and initialize a low-frequency molecular qubit composed of two states within the $J=2$ manifold, $\ket{2,-3/2,-} \equiv \ket{-3/2}$ and $\ket{2,-5/2,-} \equiv \ket{-5/2}$, with transition frequency $\approx 13.4$~kHz
(see Fig.~\ref{fig:EneLev}\textbf{b}).\\

We start the entanglement sequence by preparing the $^{40}$CaH$^+$ ion in the $\ket{-3/2}$ state in a probabilistic but heralded fashion \cite{chou_preparation_2017,choucomb2019} (also see Methods). Subsequently, we apply ground state cooling and optical pumping on the $^{40}$Ca$^+$ ion, ideally leaving the system in the state 
\begin{equation}\label{eq:psi0}
    \ket{\Psi_0}=\ket{S}\ket{-3/2}\ket{0}_M.
\end{equation}
The target entangled state of the low-frequency molecular qubit with the atom has the form
\begin{equation}\label{eq:psiL}
    \ket{\psi_L}=\frac{1}{\sqrt{2}}(\ket{S}\ket{-3/2}+\ket{D}\ket{-5/2}).
\end{equation}
This state consists of a superposition where the lower energy state of the atom and higher energy state of the molecule are paired and vice versa. The entangled state $\ket{\psi_L}$ therefore has odd parity.
Starting with $\ket{\Psi_0}$, we drive a $\pi/2$-pulse on the molecular Raman sideband transition $\ket{-3/2}\ket{0}_M \leftrightarrow \ket{-5/2}\ket{1}_M$ (see Methods) to ideally prepare
\begin{equation}\label{eq:psiI}
    \ket{\Psi_I}=\frac{1}{\sqrt{2}}\ket{S}(\ket{-3/2}\ket{0}_M+\ket{-5/2}\ket{1}_M).
\end{equation}
The intermediate state $\ket{\Psi_I}$ is an entangled state between the molecular qubit and the mode of motion $M$.
We transfer this entanglement from the motion to the $^{40}$Ca$^+$ atom by driving a $\pi$-pulse on its $\ket{S}\ket{1}_M \leftrightarrow \ket{D}\ket{0}_M$ sideband transition. This pulse of the 729~nm laser does not affect the $\ket{S}\ket{-3/2}\ket{0}_M$ component of $\ket{\Psi_I}$, while the $\ket{S}\ket{-5/2}\ket{1}_M$ component ends in the state $\ket{D}\ket{-5/2}\ket{0}_M$. In this way, the motion factors out to produce the desired entangled state $\ket{\psi_L}$ of the atom and the molecule. We start the pulse sequence on the molecule, which has zero electron spin and therefore a weak dependence of its energy levels on the external magnetic field, to reduce effects of the relatively short ($\approx$~1 ms) coherence time of the $^{40}\rm Ca^+$ qubit, which is limited by electron spin couplings to magnetic field fluctuations in our setup.\\

We characterize the entangled state with measurements of the state populations $P_{\zeta}$ (the probability of finding the atom and the molecule in the state $\ket{\zeta}$) within the four-state subspace $\{\ket{S}\ket{-5/2}$, $\ket{D}\ket{-3/2}$, $\ket{S}\ket{-3/2}$, $\ket{D}\ket{-5/2}\}$ of the atom and the molecule, and by characterizing the coherence between the states~\cite{sackett_experimental_2000}. We determine $P_{\zeta}$ by applying state-dependent fluorescence detection on the atomic states, and subsequently detecting the molecular states by transferring them to the atom via the motional mode $M$ with quantum logic, all in the same experiment trial (see Methods). Repeating the sequence of entanglement generation followed by atom and molecule state detection accumulates statistics for $P_{\zeta}$. The coherence of the entangled state can be characterized by applying
an additional ``analysis'' $\pi/2$-pulse to the atomic qubit and a $\pi/2$-pulse to the molecular qubit after the state is created, with a variable phase $\phi_a$ and $-\phi_a$, respectively, relative to the pulses used during state creation \cite{sackett_experimental_2000}. This leads to interference between the superposition parts of the entangled state, which is reflected in the populations $P_{\zeta}(\phi_a)$ observed after the $\pi/2$-pulses. In particular, the parity
\begin{equation}\label{eq:phia}
\Pi_L(\phi_a)= P_{S,-5/2}(\phi_a)+P_{D,-3/2}(\phi_a)-[P_{S,-3/2}(\phi_a)+P_{D,-5/2}(\phi_a)],  
\end{equation}
oscillates as $C \cos(2\phi_a+\phi_0)$, where $\phi_0$ is an offset in phase, and $C\geq0$ is the observed contrast. The fidelity between the entangled state produced in the experiment and $\ket{\psi_L}$ is then $F_L=\frac{1}{2}(P_{S,-3/2}+P_{D,-5/2}+C)$ (ref.\cite{sackett_experimental_2000}).\\

\begin{figure}
\centering
\includegraphics[width=\textwidth]{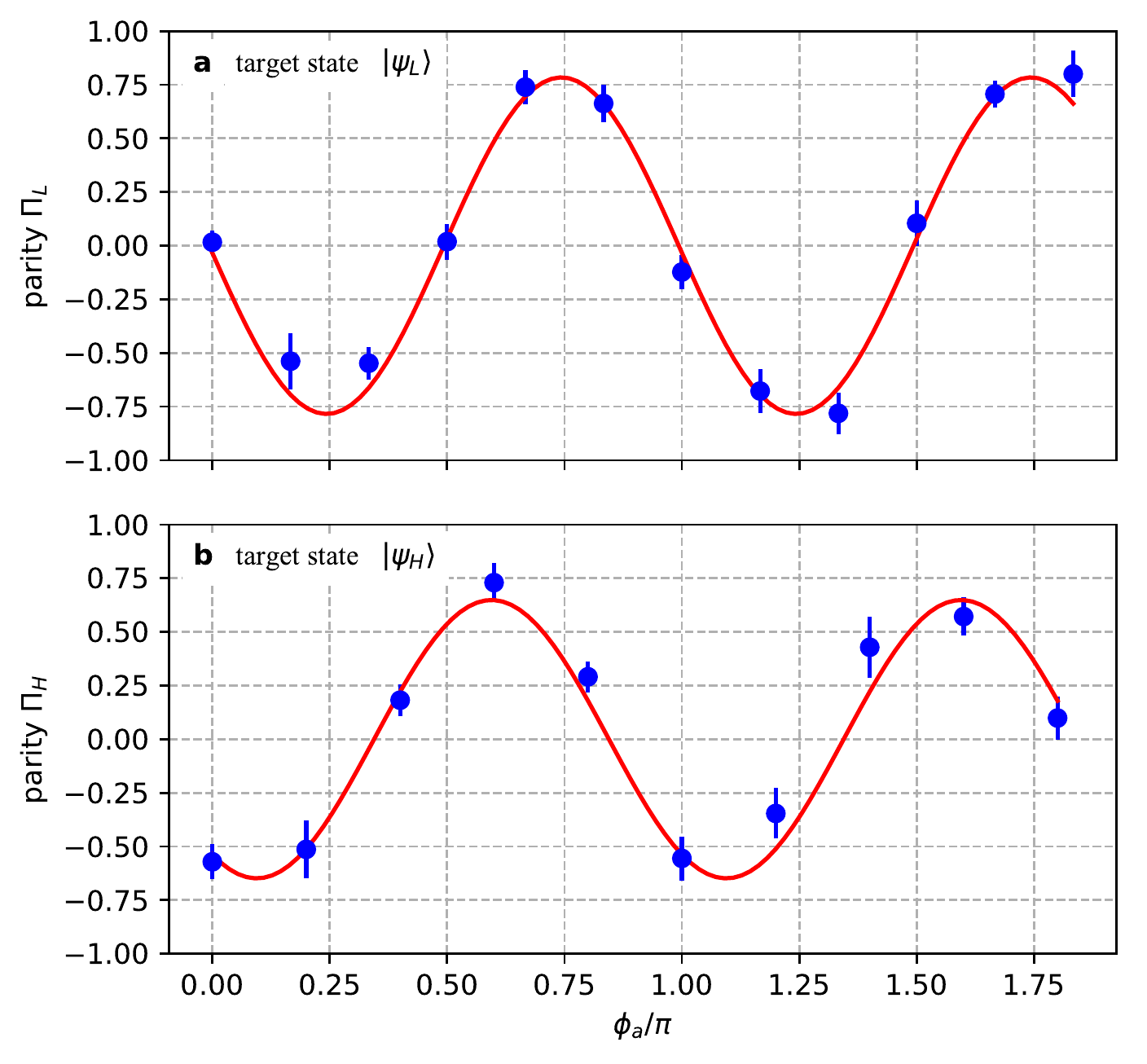}
\caption{\label{fig:ParSig} Parity measurements of the entangled states. {\bf a}, Parity fringe (see Eq.~(\ref{eq:phia})) deduced from populations after applying $\pi/2$-pulses to the experimentally prepared entangled state $\ket{\psi_L}$ of the atom and the low-frequency molecular qubit with frequency $\approx 13.4$~kHz. The phase $\phi_a$ is scanned with equal and opposite steps on the atomic and the molecular ions respectively. We observe a sinusoidal parity fringe contrast of $C=0.78(4)$ from a least-squares fit to the data, weighted by their statistical standard deviation of the mean. {\bf b}, Parity fringe (see Eq.~(\ref{eq:phiaH})), deduced from populations after applying $\pi/2$-pulses to the experimentally realized entangled state $\ket{\psi_H}$ of the atom and the high-frequency molecular qubit with frequency $\approx 855$~GHz. Here $\phi_a$ is scanned with equal steps on the two ions. We observe a parity fringe contrast of $C=0.65(5)$. The offsets of the initial phase $\phi_0$ in both figures are caused by residual Stark shifts induced by the 729~nm beam, and by the 1051~nm beams for {\bf a} and by the comb beams for {\bf b}. Error bars denote one standard deviation from the mean.
}
\end{figure}

Figure \ref{fig:ParSig}{\bf a} shows the observed parity signal (blue circles) plotted versus the analysis phase $\phi_a$, with on average $\approx$ 99 realizations of the entangled state at each phase (see Methods). The solid red line is a cosine fit of $C \cos(2 \phi_a+\phi_0)$ to the observed parity fringe, with contrast $C=0.78(4)$. Along with the populations $P_{S,-3/2}=0.50(4)$ and $P_{D,-5/2}=0.45(4)$ directly obtained from 202 experimental realizations for state $\ket{\psi_L}$ (without applying the analysis pulses), this yields 
a fidelity $F_L=0.87(3)>0.5$ indicating bipartite entanglement~\cite{leibfried_creation_2005}. This is a lower bound for the actual fidelity of the prepared state, because it also includes the infidelity introduced by imperfect readout. Additional known reductions in entangled state fidelity arise from imperfections in ground state cooling, non-ideal initial molecular state preparation into $\ket{-3/2}$, residual nonlinear coupling between different normal modes of motion, and off-resonant coupling to nearby transitions in the molecule.\\

To demonstrate the versatility of molecules, we also entangle the atom with a high-frequency molecular qubit, composed of $\ket{2} \equiv \ket{2,-3/2,-}=\ket{-3/2}$ (a state shared with the low-frequency qubit) and $ \ket{0} \equiv\ket{0,-1/2,-}$, with a transition frequency of $\approx 855$~GHz (see Fig.~\ref{fig:EneLev}\textbf{b} and Methods). The target atom-molecule entangled state in this case has the form
\begin{equation}\label{eq:psiH}
    \ket{\psi_H}=\frac{1}{\sqrt{2}}(\ket{D}\ket{2}+\ket{S}\ket{0}).
\end{equation}
The lower energy states and the higher energy states of the two ions are paired, making $\ket{\psi_H}$ an even-parity state. We manipulate the high-frequency molecular qubit using stimulated Raman transitions induced by an optical frequency comb, as theoretically described in ref.~\cite{ding_quantum_2012} and ref.~\cite{leibfried_quantum_2012} and experimentally demonstrated in ref.~\cite{choucomb2019}. The two beams originate from the same source, with frequency of each beam shifted by an acousto-optic modulator (AOM) to match the frequency differences of pairs of comb teeth with transition frequencies of the molecule to collectively drive the corresponding Raman transition (see Fig.~\ref{fig:LasBea} and Methods). After initial preparation of the system in the intermediate state $\ket{\Psi_I}$, we map $\ket{-3/2}\ket{0}_M =\ket{2}\ket{0}_M$ to $\ket{0}\ket{0}_M$ with a carrier $\pi$-pulse of the comb laser, followed by a carrier $\pi$-pulse from the 1051~nm laser that maps $\ket{-5/2}\ket{1}_M$ to $\ket{2}\ket{1}_M$. A subsequent $\ket{S}\ket{1}_M \leftrightarrow \ket{D}\ket{0}_M$ sideband $\pi$-pulse on the atom ideally prepares $\ket{\psi_H}$.\\

To quantify the fidelity $F_H$ between $\ket{\psi_H}$ and the experimentally realized entangled state with the high-frequency molecular qubit, the population measurements are accomplished in a similar way as for the entangled state involving the low-frequency molecular qubit. To find the contrast of the parity fringe we need to apply a $\pi/2$-pulse with the frequency comb to address the high-frequency molecular qubit (see Methods). By scanning the analysis phase $\phi_a$ for the  comb $\pi/2$-pulse and that of the 729~nm $\pi/2$-pulse on the atom in equal steps with the same sign, we obtain the signal shown in Fig.~\ref{fig:ParSig}\textbf{b} with
\begin{equation}\label{eq:phiaH}
\Pi_H(\phi_a)= P_{S,0}(\phi_a)+P_{D,2}(\phi_a)-[P_{S,2}(\phi_a)+P_{D,0}(\phi_a)].
\end{equation}
The signs with which the phases of the analysis $\pi/2$ pulses are scanned for the different entangled states arise from the opposite parity for the entangled components in the states $\ket{\psi_L}$ and $\ket{\psi_H}$ (see Eq.~(\ref{eq:psiL}) and (\ref{eq:psiH})), which highlights the nature of mixed-species entanglement and versatility in molecular qubit states. A fit to the parity signal (fitted contrast of 0.65(5), with on average $\approx$ 79 realizations of $\ket{\psi_H}$ per phase angle $\phi_a$) together with population measurements ($P_{S,0}=0.47(2)$ and $P_{D,2}=0.40(2)$, averaged over 491 realizations of $\ket{\psi_H}$) yields $F_H$=0.76(3). We attribute the decrease in fidelity with respect to that of $\ket{\psi_L}$ mainly to the finite coherence of the frequency comb and the larger number of imperfect operations in the pulse sequence for producing $\ket{\psi_H}$.\\

Ways to improve the entanglement fidelity include better ground state cooling and reducing the non-linear cross-coupling between motional modes \cite{roos_nonlinear_2008,nie_theory_2009}. Qubit decoherence can cause deviation of the experimentally realized state from the target pure state and decoherence of the entangled states. The leading cause of dephasing of the entangled states in our experiments is the dependence of the $^{40}\rm Ca^+$ qubit frequency on magnetic field fluctuations. For the high frequency molecular qubit, the few ms coherence time of the frequency comb leads to a significant loss in contrast of the parity curve and could be improved by several orders of magnitude with better control of the repetition rate\cite{bartels_stabilization_2004}. A cryogenic ion trap would suppress blackbody radiation and further increase the lifetime and coherence time of molecular rotational states. At the same time the vacuum would be much improved, reducing the rate of collisions of the ions with background gas that can lead to perturbation of the states and to ions trading places. Atomic qubit coherence can be improved by better stabilization of the magnetic field \cite{merkel_magnetic_2019} and by choosing a qubit that is less sensitive to magnetic field fluctuations \cite{langer_long-lived_2005}. With such improvements, higher fidelity would ensue and the coherence time of the entangled states could also be lengthened.\\

In summary, we use elements of quantum-logic spectroscopy and coherent manipulation of the resulting pure quantum states to create and characterize entanglement between long-lived electronic states of a $^{40}$Ca$^+$ atom and rotational states of a $^{40}$CaH$^+$ molecular ion, trapped together in the same potential well. We demonstrate entanglement between an atomic qubit of frequency 411.0~THz (729~nm) and molecular rotational qubits, connected by transitions of frequencies at either 13.4~kHz or 855~GHz, demonstrating the suitability of molecules for quantum state transduction between qubits of very different frequencies. We observe fidelities of the entangled states of 0.87(3) and 0.76(3), respectively. Our experimental approach is suitable for a wide range of molecular ions, offering a broad selection of qubit frequencies and properties. In particular, stimulated Raman transitions can also be driven in symmetric diatomic molecules, that have no permanent electric dipole moment and could provide qubit coherence times longer than a few minutes. This work shows that entanglement involving quantum states of a molecule is feasible and offers versatility that may be useful for a range of applications.\\

We thank Jiangfeng Du, John J. Bollinger, and Alejandra L. Collopy for carefully reading and providing feedback on this manuscript, and we thank Christoph Kurz for his help on the experimental setup. All authors contributed to planning, performing the experiments and preparation of the manuscript. This work was supported by the U. S. Army Research Office. Y. L. acknowledges support from the National Key R\&D Program of China (Grant No. 2018YFA0306600), the National Natural Science Foundation of China (Grant No. 11974330), Anhui Initiative in Quantum Information Technologies (Grant No. AHY050000).
\section{Methods}
\subsection{Atomic state manipulation and detection}

We apply laser cooling to $^{40}$Ca$^+$ only, using lasers near resonant with atomic transitions to sympathetically cool all motional modes of the two-ion crystal. We perform Doppler cooling of $^{40} \rm{Ca}^+$ with laser light at 397~nm driving the transition between $S_{1/2}$ and $P_{1/2}$ states, while using lasers at 866~nm and 854~nm to repump from the metastable $D_{3/2}$ and $D_{5/2}$ states, respectively. See Fig.~\ref{fig:LasBea} for a schematic layout of the beam lines, with blue, short-dashed arrows depicting 397~nm beams and pink, long-dashed arrows depicting 854~nm and 866~nm beams. After Doppler cooling, which cools all motional modes, we prepare the axial modes (modes parallel to the direction connecting the equilibrium positions of the two ions) near their ground state by electromagnetically-induced-transparency (EIT) cooling  \cite{roos_experimental_2000} and sideband cooling. To drive sideband transitions on the out-of-phase axial mode $M$, we apply a 729~nm pulse with a duration of 45~$\mu$s, well within the qubit coherence time of the atom, during both cooling and creating entangled states. We also apply sideband cooling on the out-of-phase radial modes \cite{roos_quantum_1999}, along two directions perpendicular to the axial mode and to each other. The out-of-phase radial modes have to be cooled close to the ground state to minimize frequency shifts and motional decoherence in the axial out-of-phase mode caused by nonlinear coupling between the radial and axial modes \cite{roos_nonlinear_2008,nie_theory_2009}. \\
\\
During fluorescence detection, we apply a near-resonant 397~nm laser to the $^{40} \rm{Ca}^+$ atom driving the transitions between levels in the $S_{1/2}$ and $P_{1/2}$ manifolds together with a 866~nm laser for repumping from the $D_{3/2}$ states. The photons scattered by the atom are directed to a photomultiplier tube (PMT) in the setup. On average, over an $85~\rm \mu s$ detection window, the PMT registers approximately 20 counts for the $\ket{S}$ state and $<0.5$ counts for the $\ket{D}$ state.

\subsection{Molecular state preparation and detection; Raman transitions and parity measurement of $\ket{\psi_H}$}
To prepare the molecule in $\ket{-3/2}$, we start with the $\ket{D}$ state for the atom, cool the motional mode $M$ close to $\ket{0}_M$ and excite it to ideally $\ket{1}_M$ with a $\pi$-pulse on the atomic sideband implementing $\ket{D}\ket{0}_M \rightarrow \ket{S}\ket{1}_M$. A $\pi$-pulse on the molecular sideband that realizes $\ket{-5/2}\ket{1}_M \rightarrow \ket{-3/2}\ket{0}_M$ and does not affect $\ket{S}\ket{0}_M$ is then attempted followed by a $\pi$-pulse on the atomic sideband $\ket{S}\ket{1}_M \rightarrow \ket{D}\ket{0}_M$. Subsequent atomic state detection would project the molecular state to $\ket{-3/2}$ if the detection outcome is $\ket{S}$. This sequence can be repeated while alternating with a sequence preparing $\ket{-5/2}$ until the confidence level for the molecular state being $\ket{-3/2}$ is above a preset threshold.
\\
\\
Detecting whether the molecule is in the $\ket{-3/2}$ state is achieved by preparing the atom and the motional mode $M$ in $\ket{D}\ket{0}_M$, applying a $\pi$-pulse on the molecular sideband $\ket{-3/2}\ket{0}_M \rightarrow \ket{-5/2}\ket{1}_M$, then a $\pi$-pulse on the atomic sideband $\ket{D}\ket{1}_M \rightarrow \ket{S}\ket{0}_M$ (which leaves $\ket{D}\ket{0}_M$ unchanged) followed by an atomic state detection. A detection outcome $\ket{S}$ indicates that the molecule was in the $\ket{-3/2}$ state. The detection outcome $\ket{D}$ is attributed to the molecule being in another molecular qubit state.\\ 
\\
To drive Raman sideband transitions and implement $\ket{\Psi_0}\rightarrow\ket{\Psi_I}$, we apply a 1051~nm laser pulse with smooth 300-$\mu$s rising and falling edges and a plateau of 162.5~$\mu$s duration, to avoid sharp pulse edges with increased frequency components driving other molecular transitions off-resonantly. \\
\\
The spectrum of the optical frequency comb has a full width at half maximum of approximately 20~nm around a center wavelength $\approx 850$~nm and a repetition rate of $f_\text{rep} \approx$ 80~MHz. For stimulated Raman transitions driven with optical frequency comb beams, we split the comb laser output into two different beams, and control their frequency and phase differences using AOMs (Fig. \ref{fig:LasBea}), with the same drive frequency $f_\text{AOM}$ but with opposite diffraction orders. The AOMs allow us to scan the frequency difference $2f_\text{AOM}$ for the beams over a range that exceeds the repetition rate. The comb is far-off resonance from any electronic transition in the molecule, but stimulated Raman transitions at frequency $f_\text{Raman}$ can be driven simultaneously by all pairs of comb teeth (one comb tooth from each beam for every pair) with matching frequency difference $f_\text{Raman}=|N f_\text{rep}-2f_\text{AOM}|$, where $N$ is an integer number of order 10000 (see below) and the sign depends on whether a photon is absorbed from the beam with $\sigma^-$ or $\pi$ polarization \cite{choucomb2019}. \\
\\
In the parity analysis for $\ket{\psi_H}$ we need to apply a $\pi/2$-pulse with the frequency comb, which has frequency components close to the $D_{5/2}\leftrightarrow P_{3/2}$ transition in $^{40}$Ca$^+$. To avoid affecting the state of the atom with this pulse, we apply sideband $\pi$-pulses on the atom mapping $\ket{D,n=0}\rightarrow\ket{S,n=1}$ to hide the atomic population in $\ket{S}$ before the comb pulse and $\ket{S,n=1}\rightarrow\ket{D,n=0}$ afterwards for the population measurements.\\

\subsection{Statistics for entangled states analyses}
To determine the fidelity for the realization of $\ket{\psi_L}$, we perform parity measurements with $\phi_a=\frac{\pi}{6}\times\{0,1,2,...,11\}$ and number of trials $\{246,39,115,106,92,83,114,62,64,67,150,50\}$, respectively. To determine the fidelity of the experimentally prepared  $\ket{\psi_H}$, we perform parity measurements with $\phi_a=\frac{\pi}{5}\times\{0,1,2,...,9\}$ and number of trials $\{98,37,132,74,141,63,52,35,84,71\}$, respectively. The variation in the number of trials is the result of repeating the experiment sequence with the same parameters including $\phi_a$ after we confirm the molecular state is heralded in the initial state of $\ket{-3/2}$. After each repetition, we check whether the molecule remains in the desired manifold of $\{\ket{-3/2},\ket{-5/2}\}$, where we design the verification measurement so that a positive result brings the molecule back to the $\ket{-3/2}$ state. If the molecule has left the $\{\ket{-3/2},\ket{-5/2}\}$ manifold we randomly draw a new value of $\phi_a$ from the list for the next time the molecular state is heralded in $\ket{-3/2}$.

\subsection{Determination of the $^{40}$CaH$^+$ $\ket{2}\equiv\ket{2,-3/2,-}\leftrightarrow\ket{0}\equiv\ket{0,-1/2,-}$ transition frequency} 
The transition frequency between the high-frequency qubit states $\ket{2,-3/2,-}$ and $\ket{0,-1/2,-}$ explored in this work is determined in a spectroscopy sequence in which the molecule is prepared in the state $\ket{2,-3/2,-}$, followed by a comb Raman pulse that transfers the population to $\ket{0,-1/2,-}$ when $|N f_\text{rep}-2f_\text{AOM}|$ is near resonance with the transition (see also ref.~\cite{choucomb2019}). The molecular population in $\ket{2,-3/2,-}$ is checked after the comb Raman pulse. Absence from the state is attributed to populating the $\ket{0,-1/2,-}$ state, because we cannot directly detect this state as in ref.~\cite{choucomb2019}. We trace out the transition lineshape, i.e. the transition probability as a function of $f_\text{AOM}$, by repeating the sequence to build up statistics while varying $f_\text{AOM}$. To determine the absolute transition frequency one needs to find the integer $N$. This is accomplished by measuring the change in frequency $\Delta f_\text{AOM}$ that needs to be applied to $f_\text{AOM}$ in order to drive the same transition when $f_\text{rep}$ is changed by $\Delta f_\text{rep}$. Specifically, to maintain $N f_\text{rep}-2f_\text{AOM} = N (f_\text{rep}+\Delta f_\text{rep})-2 (f_\text{AOM}+\Delta f_\text{AOM})$, we arrive at $N=\frac{2 \Delta f_\text{AOM}}{\Delta f_\text{rep}}$. The integer $N$ is determined in this way to be $10825$ for $f_\text{AOM}\approx 165.0$~MHz and $f_\text{rep}\approx 79.0$~MHz, and $f_\text{Raman} \approx$ 854.801 477 587~GHz. Combined with the experimentally determined rotational constant ($\approx 142.5$~GHz, ref.~\cite{choucomb2019}), this confirms that the observed transition is between the $J = 0$ and $J=2$ rotational manifolds.
\section{Data availability}
The data that support the findings of this work are available from the corresponding author upon reasonable request. 
\section{Code availability}
The computer code for analyzing the data is available from the corresponding author upon reasonable request. 
%
%

\end{document}